\documentclass[a4paper]{llncs}

\newif\ifblind
\newif\iflineno

\blindfalse
\linenofalse

\usepackage[utf8]{inputenc}
\usepackage[T1]{fontenc}
\usepackage[english]{babel}
\usepackage{amsmath}
\usepackage{amsfonts}
\usepackage{amssymb}
\usepackage{mathtools}
\usepackage{thmtools}
\usepackage{url}
\usepackage{multicol}
\usepackage{framed}
\usepackage{graphicx}
\usepackage{xcolor}
\usepackage{comment}
\usepackage{multirow}
\usepackage{etoolbox}
\usepackage{subcaption}
\usepackage{wrapfig}
\usepackage{caption}
\usepackage{bm}
\usepackage{float}
\usepackage{enumitem}
\usepackage{listings}
\usepackage{hyperref}
\usepackage{bussproofs}
\usepackage[scaled=0.85]{beramono}
\usepackage{orcidlink}

\floatstyle{plain}
\restylefloat{figure}

\setitemize{topsep=4pt}
\setenumerate{topsep=2pt}

\usepackage{tikz}
\usetikzlibrary{arrows,positioning,automata,cd}
\definecolor{LabelBlue}{RGB}{51,51,255}

\makeatletter
\newcommand{\@chapapp}{\relax}%

\newcommand*{\inlineequation}[2][]{%
  \begingroup
    \refstepcounter{equation}%
    \ifx\\#1\\\else\label{#1}\fi
    \relpenalty=10000 %
    \binoppenalty=10000 %
    \ensuremath{#2}%
    ~\@eqnnum
  \endgroup
}
\makeatother

\usepackage{array}
\newcolumntype{L}[1]{>{\raggedright\let\newline\\\arraybackslash\hspace{0pt}}m{#1}}
\newcolumntype{C}[1]{>{\centering\let\newline\\\arraybackslash\hspace{0pt}}m{#1}}
\newcolumntype{R}[1]{>{\raggedleft\let\newline\\\arraybackslash\hspace{0pt}}m{#1}}

\lstdefinelanguage{eventb}{
    morekeywords={
        END,
        CONTEXT,
        EXTENDS,
        CONSTANTS,
        SETS,
        AXIOMS,
        MACHINE,
        REFINES,
        SEES,
        VARIABLES,
        INVARIANTS,
        VARIANT,
        EVENTS,
        EVENT,
        THEN,
        ANY,
        THEORY,
        THEORIES,
        IMPORT,
        PROJECTS,
        WITH,
        WHERE,
        TYPE,
        PARAMETERS,
        AXIOMATIC,
        DEFINITIONS,
        OPERATORS,
        DATATYPES,
        DATATYPE,
        ARGS,
        DATA,
        TYPES,
        ARGUMENTS,
        constructors,
        THEOREMS,
        WHEN,
        well-definedness,
        condition,
        direct,
        definition,
        recursive,
        case,
        cases,
        PROOF,Proof,
        RULES,Rules,
        METAVARIABLES,Metavariables,
        REWRITE,Rewrite,
        REWRITES,Rewrites,
        INFERENCE,Inference,
        GIVEN,Given,
        INFER,Infer
    },
    morekeywords=[2]{
        ordinary,
        convergent,
        anticipated,
        predicate,
        expression,
        theorem,
        infix,
        commutative,
        associative
    },
    sensitive=true,
    morecomment=[l]{--},
    alsoletter=-
}

\lstdefinelanguage[]{hybrideventb}[]{eventb}{
    morekeywords={
        TIME,
        CLOCK,
        PLIANT,
        STATUS,
        INIT,
        COMPLY,
        SOLVE
    },
    morekeywords=[2]{
        pliant
    }
}

\definecolor{someGreen}{RGB}{63,127,95}
\definecolor{somePurple}{RGB}{127,0,85}
\definecolor{someBlue}{RGB}{10,80,145}

\newcommand{\POW}{\mathbb{P}}

\newcommand{\pfun}{\mathbin{\mkern6mu\mapstochar\mkern-6mu\rightarrow}}

\newcommand{\dom}{\mathrm{dom}}

\newcommand{\R}{\mathbb{R}}

\newcommand{\adv}{+}
\newcommand{\until}{\nearrow}

\newcommand{\T}{\mathbb{T}}
\newcommand{\Rp}{\R^+}          
\newcommand{\X}{\mathbb{X}}     

\newcommand{\tle}{\leq}
\newcommand{\tplus}{+}
\newcommand{\tzero}{0}
\newcommand{\tminus}{-}

\iflineno
\usepackage[left]{lineno}
\linenumbers
\fi

\title{Correct-by-Construction Design of Timed Systems in Event-B\thanks{This work was supported by ANR grant ANR-24-CE25-5742 (TAPAS project)}
}

\ifblind
\author{No author given}
\else
\author{Guillaume Dupont\,\orcidlink{0000-0002-9185-0515}\inst{1} \and Jun Sun\,\orcidlink{0000-0002-3545-1392}\inst{2}}
\institute{Toulouse INP ENSEEIHT/IRIT, \email{guillaume.dupont@toulouse-inp.fr} \and
Singapore Mangement University, \email{junsun@smu.edu.sg}}
\fi

\begin{document}

\maketitle

\begin{abstract}
Real-time systems require the careful handling of timing aspects in their models. For critical applications, this entails the use of time-aware formal methods. Currently, such formal methods only account for timing and communication layers, excluding functional aspects. Thus, they are intended to be used as \textit{a posteriori} analysis methods, on systems that have already been developed.

In contrast, methods such as Event-B have been designed to build systems incrementally using a correct-by-construction approach, but are not equipped with the ability to express timing aspects and constraints.

We propose a non-intrusive, tool-supported embedding of time and clocks in Event-B inspired by the features and semantics of timed automata. This enables the design of complex real-time systems while benefiting from the entire ecosystem and tooling support of the method. Refinement is extended to also take time into account, making it possible to design complex systems gradually in a correct-by-construction manner while integrating timing aspects from the top level.

The embedding and associated methodology are illustrated on a case study, showcasing both how timed Event-B models may be derived from timed automata, how the extended expressivity of first-order logic and set theory at the core of Event-B enables finer modelling, and how timed refinement may be used to establish complex timing properties.
\end{abstract}

\section{Introduction}\label{sec:intro}

Real-time systems designate a category of systems where timing elements (e.g., deadlines and other constraints) are central. They encompass a variety of systems, and include safety-critical applications, for which formal methods are recommended to establish important properties with the maximal degree of trust.

To that effect, pre-existing formal methods have been adjoined with \textit{time} as a top-level concept~\cite{Alur1994,Baeten1991,Davies1995,Baxter2022,Wang1998}, together with time-aware extensions of logics, typically used to encode specific properties on timed systems~\cite{Alur1994b,Maler2004,Koymans1990}.

In general, such formal methods limit their expressivity to simple timing constraints (usually, conjunctions of linear equalities with leading coefficient 1) so that verification is actually tractable, as it often involves model-checking, and thus suffers from state explosion and other usual downsides of this technique.

We also note that these formal methods' sole focus is generally on time and synchronisation/communication, making them more adapted to either high-level models of communicating real-time layers or as \textit{a posteriori} analyses conducted on pre-existing systems, to validate specific time properties, regardless of the functional behaviour of the system.

In this paper, we propose a methodology for designing real-time systems in a correct-by-construction way, based on the Event-B method augmented with timing aspects without altering the base language, through the use of \textit{theories}. The non-intrusive aspect of this extension is essential to benefit from the method's highly expressive language, its induction principle, and its rich ecosystem.

We leverage Event-B's powerful refinement operation, and extend it with timing aspects in order to be able to integrate time early during system design, and to handle it from one machine to another, specifying the system's functional behaviour in parallel with its time-related behaviour. Such operation entails timed simulation between systems, making it also suitable to correctly reduce timed systems while preserving essential properties, for model-checking purposes.

The proposed approach is illustrated on a case study that shows how it can be used to address the formalisation of a timed system, and how timed refinement may be used to express and verify timing properties on the system.

\medskip

This paper is structured as follows. Section~\ref{sec:sota} reviews existing timed formal methods and associated verification techniques. Section~\ref{sec:eventb} presents the Event-B method. Section~\ref{sec:time} defines timing aspects and lays out their integration in Event-B, while Section~\ref{sec:timeref} builds upon such definitions to define time refinement. Section~\ref{sec:cs} presents the aforementioned case study, and Section~\ref{sec:assessment} gives a comprehensive assessment of its development. Finally, Section~\ref{sec:ccl} concludes the paper.

The files associated with the case study and theories laid out in the paper have been made fully available~\cite{Dupont2026sources}.

\section{Background}\label{sec:sota}

Safety-critical real-time systems pose a number of challenges: how to model them taking into account timing and communication/synchronisation aspects, how to encode time-aware properties and how to verify them, and so on.

State-based formal methods seem to be the norm when it comes to formally model and verify timed systems: a wide variety of state-based formal methods have been extended with (dense) time. In this section, we review a few of them.

\medskip

It seems timed formal methods have existed for almost as long as non-timed formal methods. For example, time was introduced fairly early in \textit{Petri nets}, under the form of \textit{time} Petri nets~\cite{Merlin1974} and \textit{timed} Petri nets~\cite{Ramchandani1974} (TPNs), both around the same time and with the modelling and verification of asynchronous concurrent systems in mind. Both formalisms have widely been adopted nowadays, and benefit from powerful analysis tools~\cite{Berthomieu1983}.

Real Time Process Algebras~\cite{Baeten1991} have been introduced as an extension of process algebras with time progress primitives. This formalism is associated with a notion of timed (bi)simulation, although no refinement relation has been defined.

In the same vein, Hoare's \textit{Communicating Sequential Processes} formalism~\cite{Hoare1978} has been extended with timing aspects, in the form of Timed CSP~\cite{Davies1995}. This formalism is associated with time-based operational semantics~\cite{Schneider1995}, and has recently been adapted to more modern languages such as (Timed) CIRCUS~\cite{Woodcock2001,Wei2011}. Interestingly, timed CSP is associated with a specific notion of refinement~\cite{Reed1991}, that can be used to reduce models and formally check a system against its specification. It has been adorned with a formal reasoning mechanism~\cite{Dong2006}.

Similarly, \textit{tock}-CSP~\cite{Baxter2022} is close to timed CSP but relies on the \textit{tock} event for its semantics. It has been formalised in Isabelle/HOL for elaborating and proving models within the platform.

The most common timed formalism is \textit{timed automata}~\cite{Alur1994}. This formalism is central to timed formal methods as a number of them have their semantics relate to it. For example, timed CSP is semantically equivalent to (closed) timed automata~\cite{Ouaknin2002} and \textit{tock}-CSP models may be translated to timed automata~\cite{Abba2021}. The relation between time(d) Petri nets and timed automata has also been thoroughly studied~\cite{Berard2005}.

Timed automata are associated with powerful verification tools, such as  \textsc{Uppaal}~\cite{Yi1995} and \textsf{IMITATOR}~\cite{Andre2021}. Some notion of refinement has been defined~\cite{Dierks2007} but is limited to automatic verification rather than formal design.

On the Event-B side, attempts were made to address real-time systems in Event-B, for example under the form of the so-called ``active time'' approach~\cite{Cansell2006}, which has also been extended with intervals to improve time representation~\cite{Sulskus2015}. Still, such approaches do not support dense time and (dense) timed refinement, making their use limited in the context of real-time system development.

Event-B was also paired with external time-related tools~\cite{Iliasov2012,Berthing2012}; in such works however, timing aspects are effectively external, and thus disallow handling every timing aspect (including proofs) within the method.

Compared to the literature, our approach is aimed at real-time system design. It integrates time directly into Event-B and its refinement operation, in order to incorporate timing aspect as soon as possible during development.

\section{The Event-B method}\label{sec:eventb}

Event-B~\cite{Abrial2010} is a correct-by-construction, event-based formal method, aimed at the modelling of complex systems through first-order logic and set theory, making it particularly expressive. In this modelling approach, models consist of a set of \textit{components}, namely \textit{contexts} and \textit{machines}.

Contexts (Figure~\ref{lst:ebctxmch}, left) encompass the model's static data, constraints and parameters. Machines (Figure~\ref{lst:ebctxmch}, right) capture its state using variables, and its behaviour using guarded, parametrised events (Figure~\ref{lst:ebevt}), that modify these variables following a before-after predicate (BAP) relating before and after states (the latter being denoted using a prime) and denoted using the $:\mid$ notation. Among these events, the special \textit{INITIALISATION} event initialises the variables of the model, using an after-predicate (AP). Invariants are defined to capture the correctness criteria and safety of the model.

The method relies on the \textit{refinement} operation, that makes it possible to relate the behaviour of a concrete model to that of an abstract model. An Event-B development typically consists of an abstract machine, close to specification, that is refined progressively, yielding increasingly more specific machines, until the system is complete enough. Correctness is established by the user for each refinement step, ensuring the resulting machine is correct, by construction.

\begin{figure}[H]
\centering
\begin{subfigure}[c]{.47\textwidth}
\begin{subfigure}[c]{.47\textwidth}
\begin{lstlisting}[language=eventb]
CONTEXT Context
EXTENDS OtherContext
SETS $S_1$, $S_2$, ...
CONSTANTS $c_1$, $c_2$, ...
AXIOMS
 axm1: $A(c_1, c_2, \ldots)$
 ...
END
\end{lstlisting}
\end{subfigure}\hspace{8pt plus 2pt minus 2pt}%
\begin{subfigure}[c]{.48\textwidth}
\begin{lstlisting}[language=eventb]
MACHINE Machine
REFINES OtherMachine
SEES Context
VARIABLES $v_1$, $v_2$, ...
INVARIANTS
 inv1: $I(v_1, v_2,\ldots)$
 ...
EVENTS
 ...
END
\end{lstlisting}
\end{subfigure}
\caption{Event-B contexts and machines}
\label{lst:ebctxmch}
\end{subfigure}\hspace{8pt plus 2pt minus 2pt}%
\begin{subfigure}[c]{.50\textwidth}
\begin{subfigure}[c]{.43\textwidth}
\begin{lstlisting}[language=eventb]
(*\textbf{INITIALISATION}*)
THEN
 act1: $v_1,v_2,\ldots :\mid$
  $AP(v_1^\prime, v_2^\prime,\ldots)$
 ...
END
\end{lstlisting}
\end{subfigure}\hspace{8pt plus 2pt minus 2pt}%
\begin{subfigure}[c]{.52\textwidth}
\begin{lstlisting}[language=eventb]
(*\textbf{MyEvent}*)
REFINES (*\textbf{MyOtherEvent}*)
ANY $\alpha$
WHERE
 grd1: $G(\alpha,v_1,v_2,\ldots)$
 ...
THEN
 act1: $v_1,v_2 :\mid BAP(\alpha,$
  $v_1, v_1^\prime,v_2, v_2^\prime,\ldots)$
 ...
END
\end{lstlisting}
\end{subfigure}
\caption{Event-B events}
\label{lst:ebevt}
\end{subfigure}
\caption{Structure of an Event-B model}
\end{figure}

An Event-B model is associated with \textit{proof obligations} (POs) that are automatically generated and must be discharged to assert its correctness. Most notably, \textit{invariant establishment} (\ref{eqn:inv1}) ensuring the invariant is established by the initialisation, and \textit{invariant preservation} (\ref{eqn:inv2}) stating that the invariant is preserved by each event, laying out an induction principle.

\vspace{-12pt}

\begin{align}
\Gamma, AP(x^\prime) &\vdash I(x^\prime) \tag{INV1}\label{eqn:inv1} \\
\Gamma, G(\alpha,x), I(x), BAP(\alpha,x,x^{\prime}) &\vdash I(x^{\prime}) \tag{INV2}\label{eqn:inv2}
\end{align}

\vspace{-2pt}

A (correct) machine is associated with a transition system, where states correspond to the valuations of its variables and transitions correspond to event activation. The semantic of a model is thus a collection of traces where each state validates the invariant, and is linked to the next by an enabled event. These semantics enable the definition of \textit{refinement} between two machines.

\begin{figure}[ht]
\centering
\begin{subfigure}[c]{.28\textwidth}
\centering
\begin{tikzcd}[sep=large]
x^A \arrow[r, "G^A", "BAP^A"'] & x^{A\prime} \\
x^C \arrow[u, "J", -{Triangle[open]}] \arrow[r, "G^C", "BAP^C"'] & x^{C\prime} \arrow[u, "J", -{Triangle[open]}]
\end{tikzcd}
\end{subfigure}\hspace{8pt plus 2pt minus 2pt}%
\begin{subfigure}[c]{.69\textwidth}
\centering
\begin{equation}\label{eqn:gs}
\begin{split}
\Gamma, J(x^A,x^C), W(\alpha^A,\alpha^C), G^C(\alpha^C,x^C) \\ \vdash G^A(\alpha^A,x^A) 
\end{split}
\tag{GS}
\end{equation}

\vspace{-10pt}

\begin{equation}\label{eqn:sim}
\begin{split}
\Gamma, J(x^A,x^C), W(\alpha^A,\alpha^C), G^C(\alpha^C,x^C), \\ \mathit{BAP}^C(\alpha^C,x^C,x^{C\prime}) \vdash \mathit{BAP}^A(\alpha^A,x^A,x^{A\prime})
\end{split}\tag{SIM}
\end{equation}

\vspace{-6pt}
\end{subfigure}
\caption{Refinement schema and associated proof obligations}
\label{diag:ref}
\end{figure}

Refinement entails (weak) \textit{simulation} between the respective transition systems of the abstract and concrete machines~\cite{Milner1971}. The similarity of states is encoded using a \textit{gluing invariant} ($J(x^A,x^C)$) between the variables of the abstract ($A$) and concrete ($C$) machines, and via \textit{witnesses} ($W(\alpha^A,\alpha^C)$) between parameters of abstract and concrete events. Figure~\ref{diag:ref} (left) schematises this relation.

For this reason, refinement correctness is captured mainly by two proof obligations (in addition to invariant establishment/preservation): guard strengthening (\ref{eqn:gs}), stating that a concrete event can only be enabled when the refined event is, and simulation (\ref{eqn:sim}), ensuring the behaviour (BAP) of the concrete event is compatible with that of the abstract event.

To extend the expressivity of Event-B and enable the definition of reusable, high-level structures (operators, data-types), the \textit{theory} extension has been proposed and implemented within the Rodin platform~\cite{Abrial2009,Butler2013}.

This extension introduces the \textit{theory} component, which is a type-parametric set of definitions consisting of data-types, constructive and axiomatic operators, theorems and axioms, and rewriting/inference rules extending the prover.

In essence, theories make it possible to define \textit{algebraic data-types}, that can henceforth be used seamlessly in other parts of an Event-B model.

\section{Embedding time in Event-B}\label{sec:time}

To embed time in Event-B, we first gather the various elements associated with time and clocks, taken from the theory of timed automata~\cite{Alur1994}, in a set of theories. Said theories are then used to define a \textit{generic model} representing any timed system, that is refined to derive any specific timed system, in a way similar to other approaches addressing the notion of ``invited semantics''~\cite{Dupont2021,Chen2025,AitAmeur2022}.

\subsection{Basic building blocks}

We first define a number of base concepts for handling time in models. The following definitions closely follow similar elements found in timed automata theory, adapted to Event-B's expression language (first-order logic and set theory).

\paragraph{Time set}

We consider a set $\T$ of \textit{time elements} and a particular element $0 \in \T$, equipped with a commutative, associative operation $+$ for which $0$ is the neutral element. We add the property that $0$ is the only invertible element for $+$, that is: $\forall a,b \in \T, a + b = 0 \Rightarrow a = b = 0$. This operation is generalised to obtain the (finite) indexed sum, $\sum_{i = 1}^n a_i = a_1 + a_2 + \ldots + a_n$.

We then define the total order $\forall a,b \in \T, a \leq b \equiv \exists z \in \T, b = a + z$, which is compatible with $+$. For any $a,b \in \T$ with $b \leq a$, the \textit{partial} operator $-$ is uniquely defined as $z = a - b \equiv a = b + z$.

$(\T,+,\leq)$ thus defined is a cancellative, commutative, totally ordered monoid. By adding to that the \textit{least-upper-bound property}, $\T$ becomes identical in function and structure to the set of non-negative reals ($\Rp$). Time thus defined is referred to as \textit{dense}. It is the most common model of time in timed formal methods.

\begin{figure}[ht]
\centering
\begin{subfigure}{.5\textwidth}
\centering
\begin{lstlisting}[language=eventb]
THEORY TimeBase
AXIOMATIC DEFINITIONS
 TYPE $\T$
 OPERATORS
  (*0*) expression () : $\T$
  (*$+$*) expression infix ($a: \T$, $b: \T$) : $\T$
  (*$\sum$*) expression ($Ti$: $\mathbb{Z} \pfun \T$) : $\T$ -- $\color{someGreen} \sum_{i = 1}^n Ti(i)$
    well-definedness condition $\mathrm{finite}(Ti) \wedge Ti \in 1..\mathrm{card}(Ti) \rightarrow \T$
 AXIOMS
  tplus_neutral_right: $\forall a \cdot a \in \T \Rightarrow a \tplus \tzero = a$
  tplus_commutative: $\forall a,b \cdot a \in \T \wedge b \in \T \Rightarrow a \tplus b = b \tplus a$
  tplus_assoc: $\forall a,b,c \cdot a \in \T \wedge b \in \T \wedge c \in \T $
    $\Rightarrow a \tplus (b \tplus c) = (a \tplus b) \tplus c$
  tplus_eq_zero: $\forall a,b \cdot a \in \T \wedge b \in \T \wedge a \tplus b = \tzero$
    $\Rightarrow (a = \tzero \wedge b = \tzero)$
  ...
END
\end{lstlisting}
\end{subfigure}\hspace{12pt plus 2pt minus 2pt}%
\begin{subfigure}{.46\textwidth}
\centering
\begin{lstlisting}[language=eventb]
THEORY Time EXTENDS TimeBase
OPERATORS
 (*$\leq$*) predicate ($a: \T$, $b: \T$)
   direct definition $\exists z \cdot z \in \T \wedge b = a \tplus z$
AXIOMATIC DEFINITIONS
 OPERATORS
  (*$-$*) expression infix ($a: \T$, $b : \T$) : $\T$
   well-definedness condition $b \tle a$
 AXIOMS
  tle_total: $\forall a,b \cdot a \in \T \wedge b \in \T$
    $\Rightarrow (a \tle b \vee b \tle a)$
  tminus_def: $\forall a,b,z \cdot a \in \T \wedge b \in \T \wedge z \in \T$
    $\wedge b \tle a \Rightarrow (z = a \tminus b \Leftrightarrow a = b \tplus z)$
  ...
THEOREMS
 tzero_min: $\forall a \cdot a \in \T \Rightarrow 0 \tle a$
 tle_refl: $\forall a \cdot a \in \T \Rightarrow a \tle a$
 tle_antisym: $\forall a,b \cdot a \in \T \wedge b \in \T$
   $\wedge a \tle b \wedge b \tle a \Rightarrow a = b$
 ...
\end{lstlisting}
\end{subfigure}
\caption{\texttt{TimeBase} and \texttt{Time} theory excerpts}
\label{lst:timeth}
\end{figure}

Figure~\ref{lst:timeth} presents an implementation of this structure in a pair of Event-B theories. The advantage of using $\T$ over $\Rp$ (as defined for example in \cite{Su2014,Dupont2018} for Event-B) is that type-correctness and well-definedness automatically ensures any time quantity handled is non-negative. This also enables a better control on the language of constraints, and thus a better support for constraints of a certain form. Last, these theories are accompanied by a number of custom automatic and manual rewrite/inference rules, that greatly help with the proving process.

\paragraph{Clocks, valuations and constraints}

We denote the set of clocks (symbols) $\X$. A \textit{valuation} is a function $w : \X \rightarrow \T$ that maps each clock to a time value. The 0-valuation, denoted $\vec 0$ is the valuation that assigns $0$ to every clock ($\forall x \in \X, \vec 0(x) = 0$). The set of every possible valuations for clock set $\X$ is denoted $\mathcal{V}_\X$.

Given $d \in \T$ and a valuation $w \in \mathcal{V}_\X$, the valuation $(w + d)$ ($w$ advanced by $d$) is obtained by offsetting the value of $w$ by $d$ for each clock. More formally, for any $x \in \X$, $(w + d)(x) = w(x) + d$. Given a set of clocks $R \subseteq \X$ and a valuation $w \in \mathcal{V}_\X$, the valuation $[w]_R$ ($w$ with $R$ reset) is obtained by assigning $0$ to each clock $x$ in $R$, and $w(x)$ to the clocks $x$ in $\X \setminus R$.

Last, given $d\in \T$ and $w \in \mathcal{V}$, $w \nearrow d$ ($w$ until $d$) is the set of valuations offset by some value between 0 and $d$, that is: $w \nearrow d = \{ w + d' \mid d' \leq d \}$.

A \textit{constraint} $c : \X \pfun \POW(\T)$ is a (partial) function that maps each clock to a set of possible values. We say that a constraint $c$ \textit{holds for} a given valuation $w$, and we note $w \vDash c$ if and only if, for each $x \in \dom(c)$, $w(x) \in c(x)$. This notation is extended for sets of valuations: $V \vDash c \equiv \forall w \in V, w \vDash c$. This is used, for instance, for checking that a constraint holds continuously for a valuation offset up to a certain point $d$: $v \nearrow d \vDash c$. The formulation of $\vDash$ entails that clocks not mapped to a constraint by $c$ are effectively unconstrained (associated to $\top$).

This definition of constraints covers a wide variety of them. To simplify writing some common types of constraints, we define the notation $x \leq \epsilon$ ($x \in \X$, $\epsilon \in \T$) to denote the constraint where the valuation of $x$ is lower or equal to $\epsilon$. $\geq$ is defined similarly. These concepts are gathered in a theory (Figure~\ref{lst:clkth}).

\begin{figure}[ht]
\centering
\begin{subfigure}{.48\textwidth}
\centering
\begin{lstlisting}[language=eventb]
THEORY Clocks EXTENDS Time
TYPE PARAMETERS $\X$
OPERATORS
 (*$\vec 0$*) expression ()
  direct definition $(\lambda~x \cdot x \in \X \mid 0)$
 (*$+$*) expression infix ($v : \X \rightarrow \T$, $d : \T$)
  direct definition $(\lambda~x \cdot x \in \X \mid v(x) \tplus d)$
 (*$[\,]$*) expression ($v : \X \rightarrow \T$, $R : \POW(\X)$) -- $\color{someGreen} [v]_R$
  direct definition 
    $(\lambda~x \cdot x \in R \mid 0) \cup (\lambda~x \cdot x \in \X \setminus R \mid v(x))$
 (*$\nearrow$*) expression inifx ($v : \X \rightarrow \T$, $d: \T$)
  direct definition $\{ v \adv d' \mid d' \tle d \}$
 (*$\vDash$*) predicate infix ($v: \X \rightarrow \T$, $c : \X \pfun \POW(\T)$)
  direct definition $\forall x \cdot x \in \dom(c) \Rightarrow v(x) \in c(x)$
\end{lstlisting}
\end{subfigure}\hspace{12pt plus 2pt minus 2pt}%
\begin{subfigure}{.48\textwidth}
\begin{lstlisting}[language=eventb]
 (*$\vDash$*) predicate infix ($V: \POW(\X \rightarrow \T)$, $c : \X \pfun \POW(\T)$)
  direct definition $\forall v \cdot v \in V \Rightarrow v \vDash c $
 (*$\leq$*) predicate infix ($x : \X$, $e : \T$)
  direct definition $\{ x \mapsto \{ t \mid t \tle e \}\}$
 (*$\geq$*) predicate infix ($x : \X$, $e : \T$)
  direct definition $\{ x \mapsto \{ t \mid e \tle t \}\}$
THEOREMS
 advance_tzero_neutral: $\forall v \cdot v \in \X \rightarrow \T \Rightarrow v \adv \tzero = v$
 advance_advance: $\forall v,d_1,d_2 \cdot v \in \X \rightarrow \T \wedge$
  $d_1 \in \T \wedge d_2 \Rightarrow (v \adv d_1) \adv d_2 = v \adv (d_1 \tplus d_2)$
 reset_idempotent: $\forall v,R \cdot v \in \X \rightarrow \T$
   $\wedge R \subseteq \X \Rightarrow [[v]_R]_R = [v]_R$
 ...
END
\end{lstlisting}
\end{subfigure}
\caption{\texttt{Clocks} theory excerpt}
\label{lst:clkth}
\end{figure}

\vspace{-10pt}

\subsection{Timed automata}\label{ssec:ta}

Timed automata~\cite{Alur1994} are obtained by extending regular automata with time. Since it serves as the basis for our approach, it is presented succinctly in this section.

\noindent A timed automaton is a tuple $\mathcal{A} = \langle \Sigma, L, l_0, \X, I, \Xi \rangle$, where:
\begin{itemize}
\item $\Sigma$ is an alphabet of \textit{synchronisation labels};
\item $L$ is a set of \textit{locations} (or \textit{states}), and $l_0 \in L$ is the initial location;
\item $\X$ is the set of clocks;
\item $\mathcal{I} : L \rightarrow (\X \pfun \POW(\T))$ is a labelling function that maps each location to a constraint, called its \textit{invariant};
\item $\Xi$ is the set of \textit{edges}; an edge $(l,g,\sigma,R,l') \in \Xi$ consists of a source and target locations (resp. $l$ and $l' \in L$), a \textit{guard} constraint $g : \X \pfun \POW(\T)$, a label $\sigma \in \Sigma$ and a reset set $R \subseteq \X$.
\end{itemize}

The semantics of a timed automaton is given using a \textit{timed transition system} $(L\times\mathcal{V}_\X, \rightarrow_\sigma \cup \rightarrow_d)$, where each state consists of a location and a valuation, and transitions are either \textit{state} transitions ($\rightarrow_\sigma$) or \textit{time} transitions ($\rightarrow_d$). Both transition relations are given using the inference rules in Figure~\ref{fig:tasem}.

\def\defaultHypSeparation{\hskip .06in}

\begin{figure}[H]
\center
\begin{subfigure}{.58\textwidth}
\center
\small
\begin{prooftree}
\AxiomC{$(l,g,\sigma,R,l') \in \Xi$}
\AxiomC{$v \vDash g$}
\AxiomC{$[v]_R \vDash I(l')$}
\LeftLabel{state}
\TrinaryInfC{$(l,v) \rightarrow_\sigma (l',[v]_R)$}
\end{prooftree}
\end{subfigure}\hspace{12pt plus 2pt minus 2pt}%
\begin{subfigure}{.38\textwidth}
\center
\small
\begin{prooftree}
\AxiomC{$d \in \R^+$}
\AxiomC{$v \nearrow d \vDash I(l)$}
\LeftLabel{time}
\BinaryInfC{$(l,v) \rightarrow_d (l,v + d)$}
\end{prooftree}
\end{subfigure}
\caption{Timed Automata Semantics}
\label{fig:tasem}
\end{figure}

A \textit{run} of a timed automaton is a trace of that transition system with initial state $(l_0,\vec 0_\mathbb{X})$, and consisting of valid state and timed transitions.

\subsection{Generic timed system}\label{ssec:generic}

In a way similar to previous works on hybrid systems~\cite{Dupont2018}, the theories of time and clocks are used to build a generic model that represents the core features of any timed system, following the semantic rules described in Figure~\ref{fig:tasem}. Deriving a specific timed system is done by refining the generic model (\textit{instantiation}).

Figure~\ref{lst:timegen} develops the generic machine for timed system. It is assumed that a set of clocks $\X$ and a set of states $\mathit{STATES}$ are defined (in a context, typically).

\begin{figure}[ht]
\centering
\begin{subfigure}[c]{.33\textwidth}
\centering
\begin{lstlisting}[language=eventb]
MACHINE TimedGeneric
VARIABLES $state$, $v$
INVARIANTS
 inv1: $state \in \mathit{STATES}$
 inv2: $v \in \X \rightarrow \T$
EVENTS
 (*\textbf{INITIALISATION}*)
 THEN
  act1: $state,v :\mid v^\prime \in \X \rightarrow \T$
   $\wedge state^\prime \in \mathit{STATES}$
 END
\end{lstlisting}
\end{subfigure}\hspace{12pt plus 2pt minus 2pt}%
\begin{subfigure}[c]{.25\textwidth}
\centering
\begin{lstlisting}[language=eventb]
(*\textbf{TimeTransition}*)
ANY $d$, $s$, $\mathcal{I}$
WHERE
 grd1: $d \in \T$
 grd2: $state = s$
 grd3: $\mathcal{I} \in \X \pfun \POW(\T)$
 grd4: $v \until d \vDash \mathcal{I}$
THEN
 act1: $v :\mid v' = v \adv d$
END
\end{lstlisting}
\end{subfigure}\hspace{12pt plus 2pt minus 2pt}%
\begin{subfigure}[c]{.35\textwidth}
\begin{lstlisting}[language=eventb]
(*\textbf{StateTransition}*)
ANY $s$, $s_\mathit{next}$, $g$, $R$
WHERE
 grd1: $state = s$
 grd2: $s_\mathit{next} \in \mathit{STATES}$
 grd3: $g \in \X \pfun \POW(\T) \wedge v \vDash g$
 grd4: $R \subseteq \X$
THEN
 act1: $s, v :\mid s' = s_\mathit{next} \wedge v' = [v]_R$
END
\end{lstlisting}
\end{subfigure}
\caption{Timed generic model}
\label{lst:timegen}
\end{figure}

The model incorporates two aspects. On the one hand, discrete, timeless state changes are modelled using state variable $state$ and event \texttt{State\-Transition}, which is constructed similarly to an edge in a timed automaton. Time does not progress during such events, only the discrete state (clocks may be reset though).

On the other hand, the passing of time is encoded using the valuation variable $v$ and event \texttt{TimeTransition}. This event advances time by some value $d$, and is associated with a location/state $s$ and a (location) invariant $\mathcal{I}$.

Event parameters are used to make each event \textit{generic}, in the sense that they represent \textit{any} state or time transition. These parameters are to be provided using witnesses during refinement (instantiation). 

\noindent More specifically, encoding any timed automaton $\left\langle \Sigma,L,l_0,\mathbb{X},I,\Xi\right\rangle$ is done by:
\begin{enumerate}
\item Providing the invariant for each location $s \in L$: $state = s \Rightarrow v \vDash I(state)$ as machine invariants;
\item Refining \texttt{TimeTransition} for each location $s \in L$ assuming $\mathcal{I} = I(s)$;
\item Refining \texttt{StateTransition} for each edge $(s,g,\sigma,R,s_\mathit{next}) \in \Xi$.
\end{enumerate}

The use of refinement forces a certain structure in timed Event-B models. In particular, events are segregated between state and time transitions, and event involving time \textit{have to} refine other events involving time, otherwise simulation cannot be established. This is essential to enforce faithful semantics. By design, the semantics (in term of traces) of a model that follows this generic model closely abides by that of timed automata (see Figure~\ref{fig:tasem}).

\section{Timed refinement}\label{sec:timeref}

One significant characteristic of the Event-B method is its powerful refinement operation. In this section, we extend refinement by incorporating dense time, making it possible to handle timing aspects on several level of abstractions.

\paragraph{Fundamental aspects}

In term of trace semantics, refinement is characterised by a relation linking each state of the respective traces of the two machines (\textit{gluing}), and a relation linking the transitions themselves (as summarized in Figure~\ref{diag:ref}).

\begin{figure}[ht]
\centering
\begin{subfigure}{.45\textwidth}
\centering
\begin{tikzcd}[sep=normal]
(l^A,v^A) \ar[r, "g^A / \sigma^A"] & (l^{A\prime},[v^A]_{R^A}) \\
(l^C,v^C) \ar[r, "g^C / \sigma^C"] \ar[u, "\sim", -{Triangle[open]}] & (l^{C\prime},[v^C]_{R^C}) \ar[u, "\sim", -{Triangle[open]}]
\end{tikzcd}
\caption{State transition refinement}
\label{diag:tref:state}
\end{subfigure}\hspace{12pt plus 2pt minus 2pt}%
\begin{subfigure}{.45\textwidth}
\centering
\begin{tikzcd}[sep=normal]
(l^A,v^A) \ar[r, "d"] & (l^A,v^A + d) \\
(l^C,v^C) \ar[r, "d"] \ar[u, "\sim", -{Triangle[open]}] & (l^C,v^C + d) \ar[u, "\sim", -{Triangle[open]}]
\end{tikzcd}
\caption{Timed transition refinement}
\label{diag:tref:time}
\end{subfigure}
\caption{Timed refinement schema}
\label{diag:tref}
\end{figure}
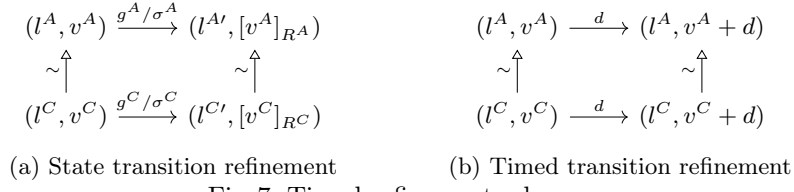

Consider two timed transition systems $(L^{A/C}\times\mathcal{V}_\X^{A/C},\rightarrow_\sigma^{A/C} \cup \rightarrow_d^{A/C})$ with superscript $A$ for \textit{abstract} and $C$ for \textit{concrete}. We suppose the existence of a gluing relation between locations and clock valuations, denoted $(l^A,v^A) \sim (l^C,v^C)$. Timed refinement is then defined following the diagrams of Figure~\ref{diag:tref}. 

State transition refinement (Figure~\ref{diag:tref:state}) is similar to the usual notion of event refinement in Event-B, as shown in Figure~\ref{diag:ref}, while timed transition refinement (Figure~\ref{diag:tref:time}) establishes a simulation relation for the timed transitions and thus links the passing of time between both systems.

These diagrams translate to two proof obligations. PO~\ref{eqn:labref} represents \textit{transition} refinement, and consists in checking that the given abstract and concrete transitions refine each other, in a way identical to the usual notion of refinement (see Figure~\ref{diag:ref}), namely guard strengthening and simulation, in addition to checking that resetting clocks preserve the gluing relation between valuations.

\vspace{-10pt}
\begin{equation}\label{eqn:labref}\tag{T-Ref}
\begin{split}
(l^C,v^C) \sim &(l^A,v^A), v^C \vDash g^C, [v^C]_{R^C} \vDash I^C(l^{C\prime}) \\
 \vdash &(l^{C\prime},[v^C]_{R^C}) \sim (l^{A\prime},[v^A]_{R^A}), v^A \vDash g^A, [v^A]_{R^A} \vDash I^A(l^{A\prime})
\end{split}
\vspace{-10pt}
\end{equation}

PO~\ref{eqn:staref} models \textit{state} or \textit{location} refinement, and corresponds to checking that timed transitions happening when the system is in a particular state/location are also simulated. Formally, it consists in ensuring that time passing maintains the gluing relationship, and that if the invariant holds in the concrete machine, it also holds in the abstract one.

\vspace{-10pt}
\begin{equation}\label{eqn:staref}\tag{S-Ref}
\begin{split}
(l^C,v^C) \sim &(l^A,v^A), v^C \nearrow d \vDash I^C(l^C) \\
\vdash & (l^C,v^C + d) \sim (l^A,v^A + d), v^A \nearrow d \vDash I^A(l^A)
\end{split}
\vspace{-10pt}
\end{equation}

\paragraph{Timed refinement in Event-B}

In the context of timed models that follow the general pattern laid out at the end of Section~\ref{ssec:generic}, we note that PO~\ref{eqn:staref} corresponds to the \textit{invariant preservation} and \textit{guard strengthening} POs of normal refinement for \texttt{TimeTransition}. Indeed, invariant preservation ensures that the simulation relation is preserved (i.e., $\Gamma \wedge v^C \sim v^A \vdash v^C + d \sim v^A + d$) while guard strengthening ensures the second part (i.e., $\Gamma \wedge v^C \nearrow d \vDash I^C(l^C) \vdash v^A \nearrow d \vDash I^A(l^A)$).

Similarly, PO~\ref{eqn:labref} corresponds to the \textit{invariant preservation} and \textit{guard strengthening} POs for the refinement of \texttt{StateTransition}: invariant preservation ensures that the locations and reset valuations remain similar and that the invariant still hold after reset, while guard strengthening ensures that the time guard in the concrete event is stronger than that in the abstract event.

This demonstrates that timed refinement is supported by Event-B without having to rely on any \textit{ad hoc} semantic extension, under the condition that models abide by the particular structure of a timed system.

\medskip

Event-B's \textit{induction principle} is a key point in our approach; it transforms a problem on entire runs/traces into sub-problems on smaller units (events). Leveraging such aspect is crucial in alleviating the proving process.

\section{Case study}\label{sec:cs}

We now illustrate the newly introduced timed features on a case study, highlighting the rich expressivity of the method, the use of the generic model and of timed refinement for verifying complex properties.

The models relative to this case studies (as well as the theories described in the previous sections) have been made available in their entirety~\cite{Dupont2026sources}.

\subsection{Presentation}\label{ssec:cspres}

This case study focuses on the modelling of a \textit{program}, i.e., a sequence of \textit{instructions}, being executed on a CPU, with the idea of matching a global execution timing with the precise timing of each instruction of the program, and then taking into account the behaviour of the CPU itself. Modelling is thus done at three levels of abstraction: the entire program as one block, the instructions of the program, and the instructions' execution on the CPU.

\paragraph{First level -- \texttt{Prog\_0}} First, we model the execution of the entire program as one block. Said execution shall happen within the $[\epsilon_\mathit{min},\epsilon_\mathit{max}]$ time interval.

\begin{figure}[ht]
\centering
\begin{tikzpicture}[font=\scriptsize]
\tikzset{->,>=stealth',node distance=3cm}
\tikzstyle{every node}=[align=center]
\tikzstyle{every state}=[minimum size=3.5em,inner sep=0]

\node[state,initial] (l0) {\textit{Init} \\ $x \leq \epsilon_\mathit{max}$} ;
\node[state,right of=l0] (l1) {\textit{Final} \\ $\top$};

\draw (l0) edge node[sloped,above]{$x \geq \epsilon_\mathit{min}$} (l1) ;

\end{tikzpicture}
\caption{Timed automaton corresponding to \texttt{Prog\_0}}
\label{fig:taprog0}
\end{figure}
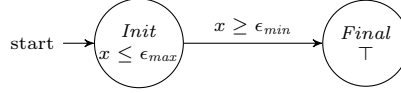

This simple behaviour is represented by the timed automaton of Figure~\ref{fig:taprog0}.

\noindent This model essentially serves as the top-level \textit{timing requirement} of the system.

\paragraph{Second level -- \texttt{Prog\_1}} Then, we model the program as an actual sequence of $n$ ($n \geq 0$) instructions each taken in instruction set $\mathit{INST}$: $P\in 1..n \rightarrow \mathit{INST}$. Each instruction is associated with a minimum and maximum execution time, $\epsilon_{i,\mathit{min}},\epsilon_{i,\mathit{max}} \in \mathit{INST} \rightarrow \T$. The current instruction being executed is identified by the \textit{program counter} $pc \in 1..n+1$. When $pc = n+1$, the program is finished.

\begin{figure}[ht]
\centering
\begin{tikzpicture}[font=\tiny]
\tikzset{->,>=stealth',node distance=3.1cm}
\tikzstyle{every node}=[align=center]
\tikzstyle{every state}=[minimum size=4em,inner sep=0]

\node[state,initial] (l0) {$pc = 1$ \\ $y \leq$ \\ $\epsilon_{i,\mathit{max}}(P(1))$} ;
\node[state,right of=l0] (l1) {$pc = 2$ \\ $y \leq$ \\ $\epsilon_{i,\mathit{max}}(P(2))$} ;
\node[right of=l1] (l2) {$\ldots$} ;
\node[state,right of=l2] (l3) {$pc = n + 1$ \\ $\top$};

\draw (l0) edge node[sloped,above]{$y \geq \epsilon_{i,\mathit{min}}(P(1))$ \\ $\{y\}$} (l1) ;
\draw (l1) edge node[sloped,above]{$y \geq \epsilon_{i,\mathit{min}}(P(2))$ \\ $\{y\}$} (l2) ;
\draw (l2) edge node[sloped,above]{$y \geq \epsilon_{i,\mathit{min}}(P(n))$ \\ $\{y\}$ } (l3) ;
\end{tikzpicture}
\caption{Timed automaton corresponding to \texttt{Prog\_1}}
\label{fig:taprog1}
\end{figure}
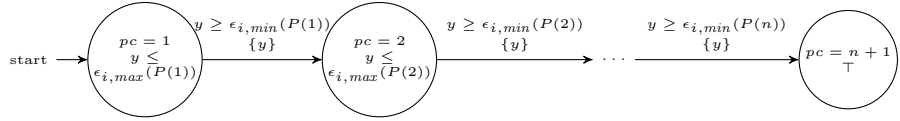

The timed automaton of this model is depicted in Figure~\ref{fig:taprog1}.

Timed refinement between this level and \texttt{Prog\_0} allows establishing that the timing constraints for each instruction from start to finish is compatible with the timing of the entire program. In this refinement, the locations $pc \in 1..n$ refine location \textit{Init}, while location $pc = n + 1$ refines \textit{Final}. Regarding transitions, every edge but the last refines ``nothing'' (\textit{skip}, in Event-B vocable), while the very last transition going into $pc = n + 1$ refines the $\mathit{Init} \rightarrow \mathit{Final}$ transition of \texttt{Prog\_0}.

\paragraph{Third level -- \texttt{Prog\_2}} Last, we model the instructions' execution on the (simplified) model of a CPU, following a fetch-exec pipeline: an instruction is queued, which takes between $\epsilon_{q,\mathit{min}}$ and $\epsilon_{q,\mathit{max}}$ time units, then decoded, which takes between $\epsilon_{d,\mathit{min}}$ and $\epsilon_{d,\mathit{max}}$, and finally executed, which takes a time that depends on the instruction being executed: $\epsilon_{x,\mathit{min}},\epsilon_{x,\mathit{max}} \in \mathit{INST} \rightarrow \T$.

\begin{figure}[ht]
\centering
\begin{subfigure}{.99\textwidth}
\centering
\begin{tikzpicture}[font=\tiny]
\tikzset{->,>=stealth',node distance=2.5cm}
\tikzstyle{every node}=[align=center]
\tikzstyle{every state}=[minimum size=4em,inner sep=0]

\node[state,initial] (l0) {$pc = 1$ \\ $q = \bot$ \\ $z \leq \epsilon_{q,\mathit{max}}$} ;
\node[state,below of=l0] (l1) {$pc = 1$ \\ $q = \top$ \\ $\top$} ;
\node[state,right of=l0] (l2) {$pc = 2$ \\ $q = \bot$ \\ $z \leq \epsilon_{q,\mathit{max}}$} ;
\node[state,below of=l2] (l3) {$pc = 2$ \\ $q = \top$ \\ $\top$} ;
\node[right of=l2] (l4) {$\ldots$} ;
\node[state,below of=l4] (l5) {$pc = n$ \\ $q = \top$ \\ $\top$} ;
\node[state,right of=l4] (l6) {$pc = n + 1$ \\ $\top$} ;

\draw (l0) edge node[sloped,above] {$z \geq \epsilon_{q,\mathit{min}}$ \\ $\{z\}$} node[sloped,below] {:queue} (l1) ;
\draw (l1) edge node[sloped,above] {:exec} (l2) ;
\draw (l2) edge node[sloped,above] {$z \geq \epsilon_{q,\mathit{min}}$ \\ $\{z\}$} node[sloped,below] {:queue} (l3) ;
\draw (l3) edge node[sloped,above] {:exec} (l4) ;
\draw (l4) edge node[sloped,above] {$z \geq \epsilon_{q,\mathit{min}}$ \\ $\{z\}$} node[sloped,below] {:queue} (l5) ;
\draw (l5) edge node[sloped,above] {:exec} (l6) ;
\end{tikzpicture}
\end{subfigure}\\[20pt]%
\begin{subfigure}{.9\textwidth}
\centering
\begin{tikzpicture}[font=\tiny]
\tikzset{->,>=stealth',node distance=2.5cm,bend angle=35}
\tikzstyle{every node}=[align=center]
\tikzstyle{every state}=[minimum size=4em,inner sep=0]

\node[state,initial] (l0) {\textit{Idle} \\ $\top$} ;
\node[state,right of=l0] (l1) {\textit{Decode} \\ $z \leq \epsilon_{d,\mathit{max}}$};
\node[state,right of=l1] (l2) {\textit{Exec} \\ $z \leq$ \\ $\epsilon_{x,\mathit{max}}(pc)$};

\draw (l0) edge node[sloped,above] {:queue} (l1) ;
\draw (l1) edge node[sloped,above] {$z \geq \epsilon_{d,\mathit{min}}$ \\ $\{z\}$} (l2) ;
\draw (l2) edge[bend left] node[sloped,below] {$z \geq \epsilon_{x,\mathit{min}}(pc)$ \\ $\{z\}$} node[sloped,above] {:exec} (l0) ;

\end{tikzpicture}
\end{subfigure}
\caption{Network of timed automata corresponding to \texttt{Prog\_2}}
\label{fig:taprog2}
\end{figure}
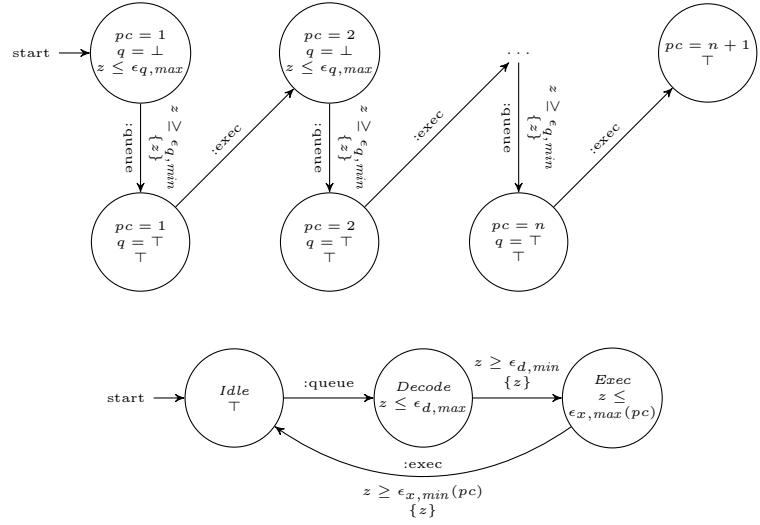

This system is modelled by the \textit{parallel composition} of two timed automata (see Figure~\ref{fig:taprog2}): one corresponding to the sequence of instructions, and the other corresponding to the executor. These two automata are synchronized on two labels: :queue (when the instruction is done being queued) and :exec (when the instruction is done being executed). We note that each instruction now has two ``inner'' states: \textit{unqueued} ($q = \bot$, before it is queued) and \textit{queued} ($q = \top$).

Timed refinement between this level and \texttt{Prog\_1} ensures that the timing aspects induced by the executor is compatible with the timing for the execution of each instruction. In this refinement, the locations $pc \in 1..n$ are refined by the locations $pc \in 1..n, q \in \{\top,\bot\}$. Studying the synchronous product reveals that locations where $q = \bot$ are coincident with location \textit{Idle} of the executor, while locations where $q = \top$ are coincident with locations \textit{Decode} and \textit{Exec}. Again, the final location ($pc = n + 1$) is refined by the final location of \texttt{Prog\_2}.

Regarding events, we notice that, for each $pc$, the transition synchronized with :exec refines the transition between $pc$ and $pc + 1$ with $pc < n$ (as they have the same post-condition). The very last transition (from $pc = n$ to $pc = n + 1$) refines the final transition of \texttt{Prog\_1}, and every other transition refines \textit{skip}.

\subsection{Realisation}

In this section, we propose a realisation of the system sketched in Section~\ref{ssec:cspres} using Event-B with timing aspects and timed refinement. The models sketched in this section have been made available in full~\cite{Dupont2026sources}.

\paragraph{Preliminaries: basic definitions and assumptions}

We begin by defining, in a context, the various objects involved in the modelling of the case study, together with a number of assumptions required for proving its correctness.

The context defines the systems' states ($\mathit{STATES} = \{\mathit{Init}, \mathit{Final}, \mathit{Idle}\}$ and $\mathit{XSTATES} = \{\mathit{Decode},\mathit{Exec}\}$), program \textit{Program} and its size $n > 0$, the (distinct) clocks $x$, $y$ and $z$ and the various timing constants used throughout models ($\epsilon_{\ast,\mathit{min}/\mathit{max}}$). For each pair of constants $\epsilon_{\ast,\mathit{min}}, \epsilon_{\ast,\mathit{max}}$, the assumption is made that the minimum constant is lower or equal to the maximum constant.

Last, timing constants have to be constrained further to be able to establish refinement correctness. Typically, because of \textit{invariant strengthening} in~\ref{eqn:labref} and~\ref{eqn:staref}, we need to ensure that the timing constants are ``compatible'' with one another. This is done using the axioms given in Figure~\ref{lst:ctx3}.

\begin{figure}[ht]
\centering
\begin{subfigure}{.9\textwidth}
\begin{lstlisting}[language=eventb]
axm14: $\epsilon_\mathit{min} \tle \sum_{k = 1}^n \epsilon_{i,\mathit{min}}(\mathit{Program}(k))$
axm15: $\sum_{k = 1}^n \epsilon_{i,\mathit{max}}(\mathit{Program}(k)) \tle \epsilon_\mathit{max}$
axm16: $\forall i \cdot i \in \mathit{INST} \Rightarrow \epsilon_{i,\mathit{min}}(i) \tle \epsilon_{q,\mathit{min}} \tplus \epsilon_{d,\mathit{min}} \tplus \epsilon_{x,\mathit{min}}(i)$
axm17: $\forall i \cdot i \in \mathit{INST} \Rightarrow \epsilon_{q,\mathit{max}} \tplus \epsilon_{d,\mathit{max}} \tplus \epsilon_{x,\mathit{max}}(i) \tle \epsilon_{i,\mathit{max}}(i)$
\end{lstlisting}
\end{subfigure}
\caption{Context -- Additional timing constraints}
\label{lst:ctx3}
\end{figure}

For instance, we need to have that $\epsilon_\mathit{min} \leq \sum_{k = 1}^n \epsilon_{i,\mathit{min}}(\mathit{Program}(k))$ (\texttt{axm14}) so that the minimal total time taken by the sequence of instructions of the program is indeed greater than the expected minimal time taken by the program as a whole.

\begin{figure}[ht]
\centering
\begin{subfigure}{.45\textwidth}
\centering
\begin{lstlisting}[language=eventb]
MACHINE Prog_0 REFINES TimedGeneric SEES Prog_Ctx
VARIABLES $state$, $v$
INVARIANTS
  saf: $state = \mathit{Init} \Rightarrow v \vDash x \leq \epsilon_\mathit{max}$
EVENTS
  (*\textbf{INITIALISATION}*)
  THEN
    act1: $state, v :\mid state' = \mathit{Init} \wedge v' = \vec 0$
  END

  (*\textbf{Init\_Location}*) REFINES (*\textbf{TimeTransition}*)
  ANY $d$
  WHERE
    grd1: $d \in \T$
    grd2: $v \until d \vDash x \leq \epsilon_\mathit{max}$
    grd3: $state = \mathit{Init}$
  WITH $s$: $s = \mathit{Init}$
       $\mathcal{I}$: $\mathcal{I} = x \leq \epsilon_\mathit{max}$
  THEN
    act1: $v :\mid v' = v \adv d$
  END
\end{lstlisting}
\end{subfigure}\hspace{12pt plus 2pt minus 2pt}%
\begin{subfigure}{.45\textwidth}
\centering
\begin{lstlisting}[language=eventb]
(*\textbf{Final\_Location}*) REFINES (*\textbf{TimeTransition}*)
ANY $d$
WHERE
  grd1: $d \in \T$
  grd2: $state = \mathit{Final}$
WITH $s$: $s = \mathit{Final}$
     $\mathcal{I}$: $\mathcal{I} = \top$ -- no invariant in Final
THEN
  act1: $v :\mid v' = v \adv d$
END

(*\textbf{End\_Program}*) REFINES (*\textbf{StateTransition}*) -- Init -> Final
WHERE
  grd1: $state = \mathit{Init}$
  grd2: $v \vDash x \geq \epsilon_\mathit{min}$
WITH $s$: $s = \mathit{Init}$, $s_\mathit{next} = \mathit{Final}$
     $g$: $g = x \geq \epsilon_\mathit{min}$
     $R$: $R = \emptyset$
THEN
  act1: $state :\mid state' = \mathit{Final}$
END
\end{lstlisting}
\end{subfigure}
\caption{Event-B machine for \texttt{Prog\_0}}
\label{lst:prog0}
\end{figure}

\paragraph{First level -- \texttt{Prog\_0}}

We now consider the modelling of the first level of abstraction. This timed system is given as a refinement of the generic model, as expected. It is depicted on Figure~\ref{lst:prog0}.

Here we find that each location (\textit{Init}, \textit{Final}) refines \texttt{TimeTransition}, while the sole transition (\texttt{End\_Program}) refines \texttt{StateTransition}. Each generic parameter is provided using witnesses.

This simple example illustrates how any timed automaton may be translated into a timed Event-B machine.

\begin{figure}[ht]
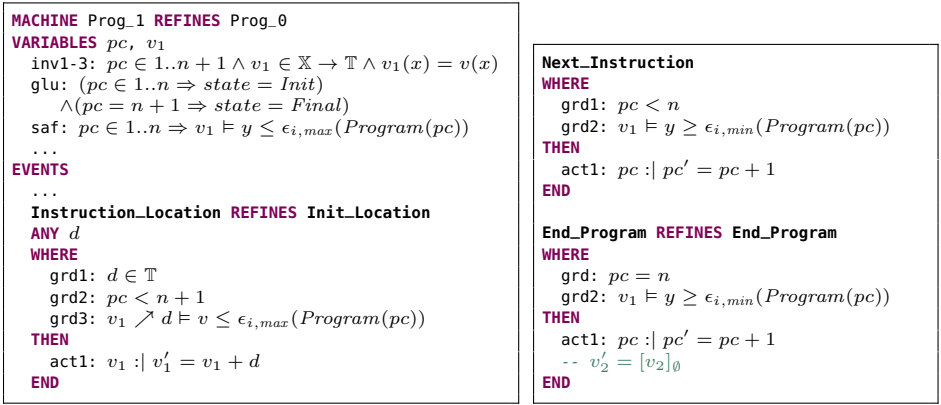

\centering
\begin{subfigure}{.54\textwidth}
\centering
\begin{lstlisting}[language=eventb]
MACHINE Prog_1 REFINES Prog_0
VARIABLES $pc$, $v_1$
  inv1-3: $pc \in 1..n+1 \wedge v_1 \in \X \rightarrow \T \wedge v_1(x) = v(x)$
  glu: $(pc \in 1..n \Rightarrow state = \mathit{Init})$
     $\wedge (pc = n + 1 \Rightarrow state = \mathit{Final})$
  saf: $pc \in 1..n \Rightarrow v_1 \vDash y \leq \epsilon_{i,\mathit{max}}(\mathit{Program}(pc))$
  ...
EVENTS
  ...
  (*\textbf{Instruction\_Location}*) REFINES (*\textbf{Init\_Location}*)
  ANY $d$
  WHERE
    grd1: $d \in \T$
    grd2: $pc < n + 1$
    grd3: $v_1 \until d \vDash v \leq \epsilon_{i,\mathit{max}}(\mathit{Program}(pc))$
  THEN
    act1: $v_1 :\mid v_1' = v_1 + d$
  END
\end{lstlisting}
\end{subfigure}\hspace{12pt plus 2pt minus 2pt}%
\begin{subfigure}{.42\textwidth}
\centering
\begin{lstlisting}[language=eventb]
(*\textbf{Next\_Instruction}*)
WHERE
  grd1: $pc < n$
  grd2: $v_1 \vDash y \geq \epsilon_{i,\mathit{min}}(\mathit{Program}(pc))$
THEN
  act1: $pc :\mid pc' = pc + 1$
END

(*\textbf{End\_Program}*) REFINES (*\textbf{End\_Program}*)
WHERE
  grd: $pc = n$
  grd2: $v_1 \vDash y \geq \epsilon_{i,\mathit{min}}(\mathit{Program}(pc))$
THEN
  act1: $pc :\mid pc' = pc + 1$
  -- $\color{someGreen} v_2' = [v_2]_\emptyset$
END
\end{lstlisting}
\end{subfigure}
\caption{Event-B machine for \texttt{Prog\_1} (excerpt)}
\label{lst:prog1}
\end{figure}

\paragraph{Second level -- \texttt{Prog\_1}} We refine machine \texttt{Prog\_0} in order to obtain \texttt{Prog\_1}. An excerpt of the machine is provided in Figure~\ref{lst:prog1}.

The machine introduces variable $pc$ modelling the program counter. Locations $pc < n + 1$ are represented by the \texttt{Instruction\_Location} event, refining \texttt{Init\_Location}, while location $pc = n + 1$ is represented by \texttt{Final\_\-Location}, refining the \texttt{Final\_\-Location} of \texttt{Prog\_0}. The link between abstract and concrete locations is established through invariant \texttt{glu}.

Transitions going from $pc$ to $pc + 1$ when $pc < n$ are represented by the \texttt{Next\_Instruction} event, and the last transition ($pc = n$ to $pc = n + 1$) is represented by \texttt{End\_Program}, which refines the \texttt{End\_Program} of \texttt{Prog\_0}.

The timed refinement POs associated with this model necessitate that we prove (in particular) that:
\begin{enumerate}
\item $\Gamma \wedge v_1 \until d \vDash y \leq \epsilon_{i,\mathit{max}}(\mathit{Program}(k)) \vdash v \until d \vDash x \leq \epsilon_\mathit{max}$ (\ref{eqn:labref}, coming from the correct refinement of \texttt{Init\_Location});
\item $\Gamma \wedge pc = n \wedge v_1 \vDash y \geq \epsilon_{i,\mathit{min}}(\mathit{Program}(pc)) \vdash v \vDash x \geq \epsilon_\mathit{min}$ (\ref{eqn:staref}, coming from the correct refinement of \texttt{End\_Program}).
\end{enumerate}

The two POs involve $y$ on the LHS and $x$ on the RHS; consequently, additional invariants bounding the evolution of $y$ and $x$ with regard to one another are needed. For instance, we notice that, when $pc < n + 1$, we necessarily have $v_1(x) \tminus v_1(y) \tle \sum_{k=1}^{pc - 1} \epsilon_{i,\mathit{max}}(\mathit{Program}(k))$ (with $v_1(x) \tminus v_1(t)$ well-defined as trivially $v_1(y) \tle v_1(x)$ is an invariant). This represents that $x$ cannot exceed the cumulative amount of time spent by $y$ in $pc < n + 1$ locations.

\paragraph{Third level -- \texttt{Prog\_2}} Last, we refine \texttt{Prog\_1} to obtain \texttt{Prog\_2}. 
In addition to $pc$, this refinement introduces variable \textit{queued} and the executor's state \textit{xstate}.

The crux of this model is to represent the synchronized parallel product. There are two aspects to it: location product, and transition product.

The analysis of Section~\ref{ssec:cspres} leads to reducing the total number of locations in the system (instead of dealing with the entire Cartesian product of states). This leaves us with four distinct state classes: $(pc < n + 1, \mathit{queued} = \bot, \mathit{xstate} = \mathit{Idle})$, $(pc < n + 1, \mathit{queued} = \top, \mathit{xstate} = \mathit{Decode})$, $(pc < n + 1, \mathit{queued} = \top, \mathit{xstate} = \mathit{Exec})$ and $(pc = n + 1, \_, \_)$. The first three location triplets refine \texttt{Instruction\_Location}, while the last one refines \texttt{Final\_Location}.

\begin{figure}[ht]
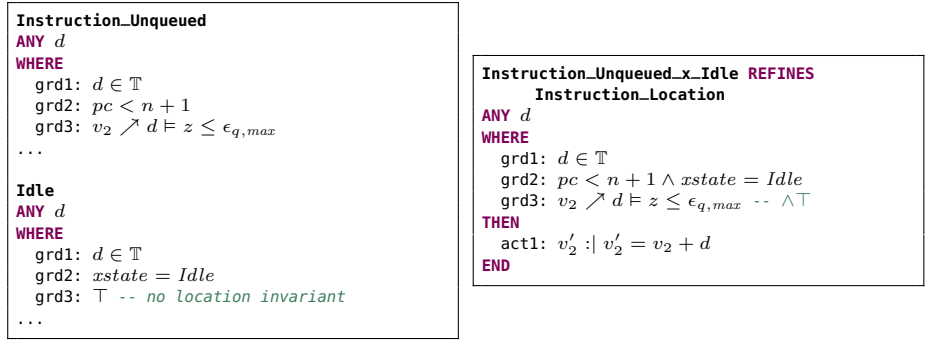

\centering
\begin{subfigure}[c]{.47\textwidth}
\centering
\begin{lstlisting}[language=eventb]
(*\textbf{Instruction\_Unqueued}*)
ANY $d$
WHERE
  grd1: $d \in \T$
  grd2: $pc < n + 1$
  grd3: $v_2 \until d \vDash z \leq \epsilon_{q,\mathit{max}}$
...

(*\textbf{Idle}*)
ANY $d$
WHERE
  grd1: $d \in \T$
  grd2: $\mathit{xstate} = \mathit{Idle}$
  grd3: $\top$ -- no location invariant
...
\end{lstlisting}
\end{subfigure}\hspace{12pt plus 2pt minus 2pt}%
\begin{subfigure}[c]{.47\textwidth}
\begin{lstlisting}[language=eventb]
(*\textbf{Instruction\_Unqueued\_x\_Idle}*) REFINES (*\textbf{Instruction\_Location}*)
ANY $d$
WHERE
  grd1: $d \in \T$
  grd2: $pc < n + 1 \wedge \mathit{xstate} = \mathit{Idle}$
  grd3: $v_2 \until d \vDash z \leq \epsilon_{q,\mathit{max}}$ -- $\color{someGreen} \wedge \top$
THEN
  act1: $v_2' :\mid v_2' = v_2 \adv d$
END
\end{lstlisting}
\end{subfigure}
\caption{Location merging}
\label{lst:prog2_1}
\end{figure}

The resulting event of each conflated state is obtained by ``merging'' the individual events of each component state, conjoining their local invariants (\texttt{grd3}) and discrete states (\texttt{grd2}). An example for the first state is provided in Figure~\ref{lst:prog2_1}.

\begin{figure}[ht]
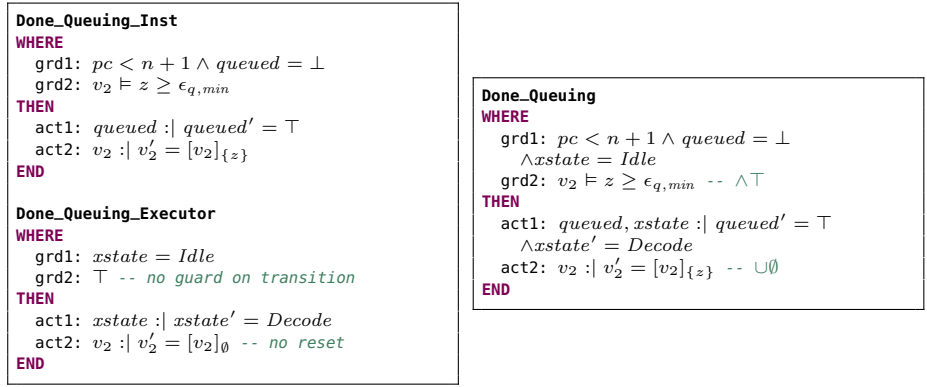

\centering
\begin{subfigure}[c]{.47\textwidth}
\centering
\begin{lstlisting}[language=eventb]
(*\textbf{Done\_Queuing\_Inst}*)
WHERE
  grd1: $pc < n + 1 \wedge \mathit{queued} = \bot$
  grd2: $v_2 \vDash z \geq \epsilon_{q,\mathit{min}}$
THEN
  act1: $\mathit{queued} :\mid \mathit{queued}' = \top$
  act2: $v_2 :\mid v_2' = [v_2]_{\{z\}}$
END

(*\textbf{Done\_Queuing\_Executor}*)
WHERE
  grd1: $\mathit{xstate} = \mathit{Idle}$
  grd2: $\top$ -- no guard on transition
THEN
  act1: $\mathit{xstate} :\mid \mathit{xstate}' = \mathit{Decode}$
  act2: $v_2 :\mid v_2' = [v_2]_\emptyset$ -- no reset
END
\end{lstlisting}
\end{subfigure}\hspace{12pt plus 2pt minus 2pt}%
\begin{subfigure}[c]{.47\textwidth}
\begin{lstlisting}[language=eventb]
(*\textbf{Done\_Queuing}*)
WHERE
  grd1: $pc < n + 1 \wedge \mathit{queued} = \bot$
    $\wedge \mathit{xstate} = \mathit{Idle}$
  grd2: $v_2 \vDash z \geq \epsilon_{q,\mathit{min}}$ -- $\color{someGreen} \wedge \top$
THEN
  act1: $\mathit{queued}, \mathit{xstate} :\mid \mathit{queued}' = \top$
    $\wedge \mathit{xstate}' = \mathit{Decode}$
  act2: $v_2 :\mid v_2' = [v_2]_{\{z\}}$ -- $\color{someGreen} \cup \emptyset$
END
\end{lstlisting}
\end{subfigure}
\caption{Transition merging}
\label{lst:prog2_2}
\end{figure}

Similarly, the components of the transitions synchronization are ``merged'' to obtain the synchronized transition event, conflating the guards and actions of each event (if they are disjoint). As an example, the synchronized product of transitions $(pc,\mathit{queued} = \bot) \rightarrow (pc, \mathit{queued} = \top) \parallel \mathit{Idle} \rightarrow \mathit{Decode}$ is given in Figure~\ref{lst:prog2_2}.

This principle is applied to $(pc,\mathit{queued} = \top) \rightarrow (pc + 1, \mathit{queued} = \bot) \parallel \mathit{Exec} \rightarrow \mathit{Idle}$, with the resulting event refining \texttt{Next\_\-Instruction}.

As for other transitions, we have $\mathit{skip} \parallel \mathit{Decode} \rightarrow \mathit{Exec} = \mathit{Decode} \rightarrow \mathit{Exec}$ refining \textit{skip}, and of course \texttt{End\_Program} refining the eponymous event of \texttt{Prog\_1}, modelling the end of the program.

Once again, refinement gives rise to the two POs~\ref{eqn:labref} and~\ref{eqn:staref}, which leads to the definition of additional invariants bounding the evolution of $z$ and $y$ with regard to one another, for each location. For instance, when $pc < n + 1$ and $\mathit{xstate} = \mathit{Exec}$, we have $v_2(y) \tminus v_2(z) \tle \epsilon_{q,\mathit{max}} \tplus \epsilon_{d,\mathit{max}}$, and so on.

\section{Assessment}\label{sec:assessment}

The presented case study showcases the expressive power of Event-B with timing aspects to handle the design and verification of a complex system, including timing properties between different levels of abstraction. This demonstrates the \textit{correct-by-construction} aspect of the Event-B method with timed refinement.

The modelling has been conducted in the Rodin platform~\cite{Abrial2010b}\footnote{\url{https://sourceforge.net/projects/rodin-b-sharp/}} and all the proof obligations have been discharged. The models have been made available in their entirety and are available for review and reuse~\cite{Dupont2026sources}.

Table~\ref{tab:pos} presents the proof statistics for the model, including the proportion of automatically vs. manually proven POs. We note a fairly high ratio of automatically proven POs, considering our intensive use of theories (which typically hinders proof automation). This is due in part to the definition of numerous proof rules, as well as to the use of sensible refinement techniques which typically alleviate and distribute some of the proving effort.

\vspace{6pt}
\begin{table}[ht]
\centering
\begin{tabular}{|l|c|c|c|}
\hline
\textbf{Model} & \# POs & \# Automatic & \# Manual\\
\hline
\texttt{TimedGeneric} & 15 & 9 (60\%) & 6 (40\%) \\
\texttt{Prog\_0} & 19 & 11 (58\%) & 8 (42\%) \\
\texttt{Prog\_1} & 51 & 25 (49\%) & 26 (51\%) \\
\texttt{Prog\_2} & 135 & 72 (53\%) & 63 (47\%) \\
\hline
\end{tabular}
\vspace{4pt}
\caption{Proof obligations associated to the development}
\label{tab:pos}
\vspace{-10pt}
\end{table}

For the most part, the manual proofs are made of a number of trivial typing and well-definedness sub-goals. The remainder typically revolves around arithmetic and inequation-related sub-goals, which are harder to prove in general. For linear inequations as used in this case study, the use of symbolic solvers could prove particularly efficient in handling this part of the reasoning.

One particular aspect missing in our approach is the capability to do animation and model-checking with our models. Indeed ProB (Rodin's model-checker/animator) does not handle Event-B theories particularly well, and especially will not be able to treat dense time properly (since it was not designed with that in mind). To address this point, timed model-checkers such as \textsc{Uppaal}~\cite{Yi1995} or \textsf{IMITATOR}~\cite{Andre2021} could be used, but would require transforming Event-B models into the formalism of each tool, and only work with specific types of constraints.

\paragraph{Comparison with model-checking-based verification tools} The main advantage of the approach is its capability to handle both timing aspects and functional aspects intertwined with one another. Timed refinement enables the gradual verification of complex timed relations between several systems, which model checkers are usually unable to achieve without \textit{ad hoc} workarounds. The expressiveness of Event-B enables devising highly general/parametrised models that can represent infinite timed automata, thus leading to general correctness results. Leveraging the method's built-in induction principle is key in order to establish complex safety and refinement properties, distributing the proof burden.

Of course, this comes with several downsides. Proof requires additional efforts, and finding the right invariants for establishing correctness is not trivial. We also note that, as for now, the verification of liveness properties is not covered by the approach, and might prove to be a significant challenge.

\section{Conclusion}\label{sec:ccl}

In this paper, we provide the theoretical and practical tools needed to extend the Event-B method with timing aspects. This leads us to propose a general methodology to implement any timed system as an Event-B machine.

Secondly, we extend Event-B's refinement to include time and leverage it to verify the consistency of timing and functional aspects of systems relative to one other. This enables the correct-by-construction design of real-time systems with time integrated early in the refinement chain.

Timed Event-B proves capable of handling real-time systems with complex timing constraints using a proof-based approach. This in turn opens the way to using all of Event-B's rich ecosystem: refinement patterns, machine (de)composition, code generation, and so on.

\medskip

As a future direction, we want to investigate the expression and verification of \textit{liveness} properties (especially reachability) expressed in formalisms such as TPTL~\cite{Alur1994b}, TCTL~\cite{Alur1993} or MTL~\cite{Koymans1990}. This type of properties is typically difficult to encode and prove in Event-B, although recent works~\cite{Ferrarotti2024,Riviere2023} has laid out promising approaches in this direction.

Multi-clock and discrete clock systems and their refinement ought to be investigated, for handling cyber-physical aspects and bringing models closer to implementations. On that topic, continuous refinement has already been investigated for hybrid systems~\cite{Dupont2019}, and may be adapted to timed Event-B.

Lastly, the link between Rodin and timed model-checkers must be investigated, in order to simulate timed Event-B models, and even help with proofs and constraint synthesis. Used in the other direction, Event-B could be used to reduce complex timed systems in order to make model-checking tractable while preserving certain established properties.

\bibliographystyle{splncs03}
\bibliography{biblio}

@techreport{Milner1971,
    title        = {An Algebraic Definition of Simulation Between Programs},
    author       = {Milner, Robin},
    institution  = {Computer Science Department, Standford University},
    year         = 1971,
    address      = {Stanford, CA, USA}
}

@article{Baeten1991,
    title={Real time process algebra},
    author={Baeten, J. C. M. and Bergstra, J. A.},
    year={1991},
    journal={Formal Aspects of Computing},
    pages={142--188},
    volume={3},
    issue={2},
    doi={10.1007/BF01898401}

}

@article{Davies1995,
title = {A brief history of {T}imed {CSP}},
journal = {Theoretical Computer Science},
volume = {138},
number = {2},
pages = {243-271},
year = {1995},
note = {Meeting on the mathematical foundation of programing semantics},
issn = {0304-3975},
doi = {https://doi.org/10.1016/0304-3975(94)00169-J},
author = {Jim Davies and Steve Schneider}
}

@article{Schneider1995,
title = {An Operational Semantics for {T}imed {CSP}},
journal = {Information and Computation},
volume = {116},
number = {2},
pages = {193-213},
year = {1995},
issn = {0890-5401},
doi = {https://doi.org/10.1006/inco.1995.1014},
author = {S. Schneider}
}

@InProceedings{Reed1991,
author="Reed, G. M.
and Roscoe, A. W.
and Schneider, S. A.",
editor="Morris, Joseph M.
and Shaw, Roger C.",
title="{CSP} and Timewise Refinement",
booktitle="4th Refinement Workshop",
year="1991",
publisher="Springer London",
address="London",
pages="258--280"
}

@inproceedings{Wei2011,
  author={Wei, Kun and Woodcock, Jim and Burns, Alan},
  booktitle={2011 16th IEEE International Conference on Engineering of Complex Computer Systems}, 
  title={Timed {C}ircus: {T}imed {CSP} with the Miracle}, 
  year={2011},
  volume={},
  number={},
  pages={55-64},
  doi={10.1109/ICECCS.2011.13}
}

@article{Ouaknin2002,
title = {Timed {CSP} = Closed Timed Automata},
journal = {Electronic Notes in Theoretical Computer Science},
volume = {68},
number = {2},
pages = {142-159},
year = {2002},
note = {EXPRESS'02, 9th International Workshop on Expressiveness in Concurrency (Satellite Workshop of CONCUR 2002)},
issn = {1571-0661},
doi = {https://doi.org/10.1016/S1571-0661(05)80369-X},
author = {Joël Ouaknine and James Worrell}
}

@inproceedings{Dong2006,
  title={A reasoning method for timed {CSP} based on constraint solving},
  author={Dong, Jin Song and Hao, Ping and Sun, Jun and Zhang, Xian},
  booktitle={International Conference on Formal Engineering Methods},
  pages={342--359},
  year={2006},
  organization={Springer}
}

@article{Baxter2022,
    title={Sound reasoning in tock-{CSP}},
    author={Baxter, James and Ribeiro, Pedro and Cavalcanti, Ana},
    journal={Acta Informatica},
    year={2022},
    pages={125--162},
    volume={59},
    issue={1},
    doi={10.1007/s00236-020-00394-3}
}

@Inbook{Wang1998,
author="Wang, Jiacun",
title="Time Petri Nets",
bookTitle="Timed {P}etri Nets: Theory and Application",
year="1998",
publisher="Springer US",
address="Boston, MA",
pages="63--123",
isbn="978-1-4615-5537-7",
doi="10.1007/978-1-4615-5537-7_4"
}

@InProceedings{Abba2021,
author="Abba, Abdulrazaq
and Cavalcanti, Ana
and Jacob, Jeremy",
editor="Campos, S{\'e}rgio
and Minea, Marius",
title="Temporal Reasoning Through Automatic Translation of tock-{CSP} into Timed Automata",
booktitle="Formal Methods: Foundations and Applications",
year="2021",
publisher="Springer International Publishing",
address="Cham",
pages="70--86"
}

@InProceedings{Maler2004,
author="Maler, Oded
and Nickovic, Dejan",
editor="Lakhnech, Yassine
and Yovine, Sergio",
title="Monitoring Temporal Properties of Continuous Signals",
booktitle="Formal Techniques, Modelling and Analysis of Timed and Fault-Tolerant Systems",
year="2004",
publisher="Springer Berlin Heidelberg",
address="Berlin, Heidelberg",
pages="152--166"
}

@article{Koymans1990,
    author={Koymans, Ron},
    title={Specifyin real-time properties with metric temporal logigc},
    journal={Real-Time Systems},
    year={1990},
    pages="255--299",
    volume={2},
    issue={4},
    doi={10.1007/BF01995674}
}

@article{Alur1994b,
author = {Alur, Rajeev and Henzinger, Thomas A.},
title = {A really temporal logic},
year = {1994},
issue_date = {Jan. 1994},
publisher = {Association for Computing Machinery},
address = {New York, NY, USA},
volume = {41},
number = {1},
issn = {0004-5411},
url = {https://doi.org/10.1145/174644.174651},
doi = {10.1145/174644.174651},
journal = {J. ACM},
month = jan,
pages = {181–203},
numpages = {23}
}

@article{Alur1993,
title = {Model-Checking in Dense Real-Time},
journal = {Information and Computation},
volume = {104},
number = {1},
pages = {2-34},
year = {1993},
issn = {0890-5401},
doi = {10.1006/inco.1993.1024},
author = {R. Alur and C. Courcoubetis and D. Dill}
}

@book{Merlin1974,
  title={A Study of the Recoverability of Computing Systems.},
  author={Merlin, Philip Meir},
  year={1974},
  publisher={University of California, Irvine}
}

@book{Ramchandani1974,
  title={Analysis of asynchronous concurrent systems by timed {P}etri nets},
  author={Ramchandani, Chander},
  year={1974},
  publisher={Massachusetts Institute of Technology}
}

@inproceedings{Berthomieu1983,
  TITLE = {An enumerative approach for analyzing Time {P}etri Nets},
  AUTHOR = {Berthomieu, Bernard and Menasche, Miguel},
  URL = {https://laas.hal.science/hal-04187100},
  BOOKTITLE = {{IFIP 9th World Computer Congress}},
  ADDRESS = {Paris, France},
  YEAR = {1983},
  MONTH = Sep
}

@article{Hoare1978,
author = {Hoare, C. A. R.},
title = {Communicating sequential processes},
year = {1978},
issue_date = {Aug. 1978},
publisher = {Association for Computing Machinery},
address = {New York, NY, USA},
volume = {21},
number = {8},
issn = {0001-0782},
doi = {10.1145/359576.359585},
journal = {Commun. ACM},
month = aug,
pages = {666–677},
numpages = {12}
}

@inproceedings{Woodcock2001,
  title={A concurrent language for refinement},
  author={Woodcock, Jim and Cavalcanti, Ana},
  booktitle={5th Irish workshop on formal methods},
  year={2001},
  organization={BCS Learning \& Development}
}

@InProceedings{Berard2005,
author="B{\'e}rard, Beatrice
and Cassez, Franck
and Haddad, Serge
and Lime, Didier
and Roux, Olivier H.",
editor="Pettersson, Paul
and Yi, Wang",
title="Comparison of the Expressiveness of Timed Automata and Time {P}etri Nets",
booktitle="Formal Modeling and Analysis of Timed Systems",
year="2005",
publisher="Springer Berlin Heidelberg",
address="Berlin, Heidelberg",
pages="211--225"
}

@book{Abrial2010,
    title        = {Modeling in {Event-B}: {S}ystem and Software Engineering},
    author       = {Abrial, Jean-Raymond},
    year         = 2010,
    publisher    = {Cambridge University Press},
    address      = {New York, NY, USA},
    isbn         = 9781139637794,
    edition      = {1st}
}

@techreport{Abrial2009,
    title        = {Proposals for Mathematical Extensions for {Event-B}},
    author       = {Abrial, Jean-Raymond and Butler, Michael and Hallerstede, Stefan and Leuschel, Michael and Schmalz, Matthias and Voisin, Laurent},
    year         = 2009,
    institution  = {(No Institution)}
}

@incollection{Butler2013,
    title        = {Practical Theory Extension in {Event-B}},
    author       = {Butler, Michael and Maamria, Issam},
    year         = 2013,
    booktitle    = {Theories of Programming and Formal Methods},
    publisher    = {Springer Berlin Heidelberg},
    series       = {Lecture Notes in Computer Science},
    volume       = 8051,
    pages        = {67--81},
    isbn         = {978-3-642-39697-7},
    editor       = {Liu, Zhiming and Woodcock, Jim and Zhu, Huibiao}
}

@article{Abrial2010b,
    title = {Rodin: an open toulset for modelling and reasoning in {E}vent-{B}},
    author = {Abrial, Jean-Raymond and Butler, Michael and Hallerstede, Stefan and Hoang, Thai Son and Mehta, Farhad and Voisin, Laurent},
    journal = {International Journal on Software Tools for Technology Transfer},
    year={2010},
    pages={447-466},
    volume={12},
    issue={6}
}

@article{Su2014,
    title        = {Formalizing hybrid systems with {Event-B} and the {R}odin Platform},
    author       = {Su, Wen and Abrial, Jean-Raymond and Zhu, Huibiao},
    year         = 2014,
    journal      = {Science of Computer Programming},
    volume       = {94, Part 2},
    pages        = {164--202},
    issn         = {0167-6423},
    note         = {Abstract State Machines, Alloy, B, {VDM}, and Z Selected and extended papers from {ABZ} 2012}
}

@inproceedings{Dupont2018,
    title        = {Proof-Based Approach to Hybrid Systems Development: Dynamic Logic and {E}vent-{B}},
    author       = {Dupont, Guillaume and A{\"i}t-Ameur, Yamine and Pantel, Marc and Singh, Neeraj Kumar},
    year         = 2018,
    booktitle    = {Abstract State Machines, Alloy, B, TLA, VDM, and Z},
    publisher    = {Springer International Publishing},
    address      = {Cham},
    pages        = {155--170},
    doi          = {10.1007/978-3-319-91271-4_11},
    editor       = {Butler, Michael and Raschke, Alexander and Hoang, Thai Son and Reichl, Klaus}
}

@inproceedings{Dupont2019,
    title        = {Handling Refinement of Continuous Behaviors: A Refinement and Proof Based Approach with {E}vent-{B}},
    author       = {Dupont, Guillaume and A{\"i}t-Ameur, Yamine and Pantel, Marc and Singh, Neeraj K.},
    year         = 2019,
    booktitle    = {13th International Symposium TASE},
    publisher    = {IEEE Computer Society Press},
    pages        = {9--16}
}

@article{Dupont2021,
    author  = {Dupont, Guillaume and A{\"i}t-Ameur, Yamine and Singh, Neeraj K. and Pantel, Marc},
    title   = {{E}vent-{B} Hybridation: {A} Proof and Refinement-based Framework for Modelling Hybrid Systems},
    journal = {ACM Transactions on Embedded Computing Systems},
    volume  = {20},
    number  = {4},
    doi     = {https://doi.org/10.1145/3448270},
    year    = {2021}
}

@article{Alur1994,
title = {A theory of timed automata},
journal = {Theoretical Computer Science},
volume = {126},
number = {2},
pages = {183-235},
year = {1994},
issn = {0304-3975},
doi = {https://doi.org/10.1016/0304-3975(94)90010-8},
url = {https://www.sciencedirect.com/science/article/pii/0304397594900108},
author = {Rajeev Alur and David L. Dill},
}

@Inbook{Yi1995,
author="Yi, Wang
and Pettersson, Paul
and Daniels, Mats",
title="Automatic Verification of Real-Time Communicating Systems by Constraint-Solving",
bookTitle="Formal Description Techniques VII: Proceedings of the 7th IFIP WG 6.1 international conference on formal description techniques",
year="1995",
publisher="Springer US",
address="Boston, MA",
pages="243--258",
isbn="978-0-387-34878-0",
doi="10.1007/978-0-387-34878-0_18",
url="https://doi.org/10.1007/978-0-387-34878-0_18"
}

@InProceedings{AitAmeur2022,
author="A{\"i}t-Ameur, Yamine
and Dupont, Guillaume
and Mendil, Ismail
and M{\'e}ry, Dominique
and Pantel, Marc
and Rivi{\`e}re, Peter
and Singh, Neeraj K.",
editor="ter Beek, Maurice H.
and Monahan, Rosemary",
title="Empowering the {E}vent-{B} Method Using External Theories",
booktitle="Integrated Formal Methods",
year="2022",
publisher="Springer International Publishing",
address="Cham",
pages="18--35"
}

@InProceedings{Ferrarotti2024,
author="Ferrarotti, Flavio
and Rivi{\`e}re, Peter
and Schewe, Klaus-Dieter
and Singh, Neeraj Kumar
and Ameur, Yamine A{\"i}t",
editor="Meier, Arne
and Ortiz, Magdalena",
title="A Complete Fragment of {LTL(EB)}",
booktitle="Foundations of Information and Knowledge Systems",
year="2024",
publisher="Springer Nature Switzerland",
address="Cham",
pages="237--255"
}

@InProceedings{Riviere2023,
author="Rivi{\`e}re, P.
and Singh, N. K.
and A{\"i}t-Ameur, Y.
and Dupont, G.",
editor="Rozier, Kristin Yvonne
and Chaudhuri, Swarat",
title="Formalising Liveness Properties in {E}vent-{B} with the Reflexive {EB4EB} Framework",
booktitle="NASA Formal Methods",
year="2023",
publisher="Springer Nature Switzerland",
address="Cham",
pages="312--331"
}

@inproceedings{Chen2025,
  author       = {Christophe Chen and
                  Peter Rivi{\`{e}}re and
                  Neeraj Kumar Singh and
                  Guillaume Dupont and
                  Yamine A{\"{\i}}t{-}Ameur and
                  Marc Frappier},
  title        = {A Proof-Based Ground Algebraic Meta-Model for Reasoning on {ASTD}
                  in {E}vent-{B}},
  booktitle    = {13th {IEEE/ACM} International Conference on Formal Methods in Software
                  Engineering, FormaliSE@ICSE 2025, Ottawa, ON, Canada, April 27-28,
                  2025},
  pages        = {46--57},
  publisher    = {{IEEE}},
  year         = {2025},
  doi          = {10.1109/FORMALISE66629.2025.00011}
}

@InProceedings{Andre2021,
author="Andr{\'e}, {\'E}tienne",
editor="Silva, Alexandra
and Leino, K. Rustan M.",
title="{IMITATOR} 3: {S}ynthesis of Timing Parameters Beyond Decidability",
booktitle="Computer Aided Verification",
year="2021",
publisher="Springer International Publishing",
address="Cham",
pages="552--565"
}

@InProceedings{Dierks2007,
author="Dierks, Henning
and Kupferschmid, Sebastian
and Larsen, Kim G.",
editor="Raskin, Jean-Fran{\c{c}}ois
and Thiagarajan, P. S.",
title="Automatic Abstraction Refinement for Timed Automata ",
booktitle="Formal Modeling and Analysis of Timed Systems",
year="2007",
publisher="Springer Berlin Heidelberg",
address="Berlin, Heidelberg",
pages="114--129",
doi="10.1007/978-3-540-75454-1_10"
}

@InProceedings{Cansell2006,
author="Cansell, Dominique
and M{\'e}ry, Dominique
and Rehm, Joris",
editor="Julliand, Jacques
and Kouchnarenko, Olga",
title="Time Constraint Patterns for Event B Development",
booktitle="B 2007: Formal Specification and Development in B",
year="2006",
publisher="Springer Berlin Heidelberg",
address="Berlin, Heidelberg",
pages="140--154",
doi="10.1007/11955757_13"
}

@InProceedings{Sulskus2015,
author="Sulskus, Gintautas
and Poppleton, Michael
and Rezazadeh, Abdolbaghi",
editor="Dastani, Mehdi
and Sirjani, Marjan",
title="An Interval-Based Approach to Modelling Time in Event-B",
booktitle="Fundamentals of Software Engineering",
year="2015",
publisher="Springer International Publishing",
address="Cham",
pages="292--307",
doi="10.1007/978-3-319-24644-4_20"
}

@INPROCEEDINGS{Iliasov2012,
  author={Iliasov, Alexei and Romanovsky, Alexander and Laibinis, Linas and Troubitsyna, Elena and Latvala, Timo},
  booktitle={2012 First International Workshop on Formal Methods in Software Engineering: Rigorous and Agile Approaches (FormSERA)}, 
  title={Augmenting Event-B modelling with real-time verification}, 
  year={2012},
  volume={},
  number={},
  pages={51-57},
  doi={10.1109/FormSERA.2012.6229789}
}

@InProceedings{Berthing2012,
author="Berthing, Jesper
and Bostr{\"o}m, Pontus
and Sere, Kaisa
and Tsiopoulos, Leonidas
and Vain, J{\"u}ri",
editor="Derrick, John
and Gnesi, Stefania
and Latella, Diego
and Treharne, Helen",
title="Refinement-Based Development of Timed Systems",
booktitle="Integrated Formal Methods",
year="2012",
publisher="Springer Berlin Heidelberg",
address="Berlin, Heidelberg",
pages="69--83",
doi="10.1007/978-3-642-30729-4_6"
}

@misc{Dupont2026sources,
  author       = {Dupont, Guillaume and
                  Sun, Jun},
  title        = {Correct-by-construction Design of Timed System in
                   Event-B: Case Study
                  },
  month        = {may},
  year         = {2026},
  publisher    = {Zenodo},
  version      = {1.0},
  doi          = {10.5281/zenodo.20465023},
  url          = {https://doi.org/10.5281/zenodo.20465023},
}

\end{document}